\documentclass[11pt]{article}

\usepackage{graphicx}
\usepackage{amssymb}
\usepackage[english,francais]{babel}
\RequirePackage[latin1]{inputenc}  
	\usepackage{vmargin}
	\setpapersize{A4}
	\setmarginsrb{2cm}{2cm}{2cm}{2cm}{1cm}{1cm}{0cm}{1.5cm}

	\newcommand {\indrm}[1] {_\mathrm{#1}}
	\newcommand {\exprm}[1] {^\mathrm{#1}}
	\newcommand {\vect}[1] {\overrightarrow{#1}}
	\newcommand {\tens}[1] {\underline{#1}}
	 \newcommand {\TENS}[1] {\mathbb{#1}}
	
	\newcommand {\trace}{\mathrm{tr}}

		\newcommand {\SIG} {\tens{\sigma}}
		\newcommand {\SIGD} {\tens{\sigma} \exprm{d}}
		
		\newcommand {\si} {\sigma \indrm {I}}
		\newcommand {\sii} {\sigma \indrm {II}}
		\newcommand {\siii} {\sigma \indrm {III}}

		\newcommand {\so} {\sigma \indrm 0}
		\newcommand {\sy} {\sigma \indrm y}
		\newcommand {\sic} {\sigma \indrm c}
		\newcommand {\sit} {\sigma \indrm t}
		
		\newcommand{\PD} {\TENS{P}^{\mathrm{d}}}
		\newcommand {\A} {\tens{A}}
		\newcommand {\B} {\tens{B}}
		\newcommand{\ti}{\mathrm{i}}
		\newcommand{\tj}{\mathrm{j}}
		\newcommand{\tI}{\mathrm{I}}
		\newcommand{\tJ}{\mathrm{J}}
		\newcommand{\tk}{\mathrm{k}}
		\newcommand{\tl}{\mathrm{l}}
		\newcommand{\tp}{\mathrm{p}}
		\newcommand{\tq}{\mathrm{q}}
		\newcommand{\tr}{\mathrm{r}}
		\newcommand{\ts}{\mathrm{s}}
		\newcommand{\pei}{p_{\mathrm{i}}}
		\newcommand{\pej}{p_{\mathrm{j}}}

\begin{document}

\title{A new yield criterion for the concrete materials}
\author{Marc François} 
\date{}
\maketitle

\begin{center}
Cr. Acad. Sci. IIb 336, p. 417-421 (2008)
\end{center}

\noindent{\bf Abstract}
\vskip 0.5\baselineskip
\noindent
The proposed yield criterion depends upon two material constants and is proven to be smooth and convex under a simple condition. These properties induce a mathematical robustness that allows a further use in a damage mechanics model. The analytical gradient and hessian are given. The obtained yield surface is relevant to Kupfer's biaxial testings on concrete. The identification procedure, with respect to the classical uniaxial tension and compression testings, is detailed.

\vskip 0.5\baselineskip

\selectlanguage{francais}
\noindent{\bf R\'esum\'e}
\vskip 0.5\baselineskip
\noindent
{\bf Un nouveau critère de limite d'élasticité pour les bétons. }
Le critère proposé utilise deux constantes matériaux. Son gradient et son hessien sont donnés, sa convexité est démontrée, cette dernière ne dépendant que de la positivité des constantes matériau. Ces propriétés induisent une robustesse mathématique qui autorise son emploi futur au sein d'un modèle d'endommagement. La surface de charge obtenue est en bonne adéquation avec les essais biaxiaux de Kupfer. La procédure d'identification, à partir des contraintes limites d'élasticité identifiées en traction et en compression, est donnée. 

keyword damage; yield function; concrete
\vskip 0.5\baselineskip
\noindent{\small{\it Mots-cl\'es~:} endommagement~; critère~;
béton}


\newpage

\selectlanguage{english}
\section{Introduction}
\label{intro}
Kupfer \cite{kupfer_69} performed biaxial testings on concrete ($\si, \sii, \siii=0$ in principal stresses) that are recognized as a reference and relevant for the stress states met in classical civil engineering. Although his best known results are the peak stress curves, the figure 15 of this article exhibits the elastic domain of concrete and the present work is referred to it. Kupfer measured the elastic limit from the loss of linearity of the stress to strain curve: this identification depends upon the precision of the measure. Some authors obtain, in pure tension, a yield stress very close to the peak stress \emph{i.e.} a fragile behavior; on the contrary, Kupfer identified the yield stress in tension approximately as the half of the peak stress. A similar result is also obtained by Terrien \cite{terrien_80} thanks to a very accurate method, confirming Kupfer's results.\\
Most of old elasticity criteria, such as Mohr-Coulomb's one, or more recent approaches \cite{Hoek_80}, are based on stress vector considerations. They may be accurate, have a strong mechanical sense and allow to define the crack orientation, but are non smooth. These corners are not experimental evidences and lead to numerical difficulties when the criterion is used in a damage model.\\
Actually, the most used criteria are simple functions of the stress tensor $\SIG$ and the stress deviator $\SIGD$ classical invariants $I_1=\trace(\SIG)$, $J_2'=\trace(\SIGD.\SIGD)/2$ and $J_3'=\trace(\SIGD.\SIGD.\SIGD)/3$ whose polynomial structure insure the smoothness of the yield surface. Among them, we shall compare to the present model the De Vree \cite{devree_93}, Willam \cite{etse_94} and Maïolino \cite{maiolino_05} expressions (all using two constants). The identification will be made preferentially with respect to the uniaxial compression and tension yield stresses, respectively denoted by points C and T on figure (\ref{YSfig}).\\
 The De Vree criterion depends upon $(I_1,J_2')$; its identification, with respect to the points C and T, gives the constants $(\gamma=7.75, \kappa=3.9\,10^{-5})$. 
The Maïolino's criterion has been recently proposed for rock materials; it takes into account the effect of the third invariant $J_3'$ and, in order to keep a two-constants form, a linear dependancy $(\sigma^+=I_{1m}-I_{1})$ has been retained here. The model, in this form, cannot fit both points C and T (the identified constant $L_{s}=0,5$ reaches the bound allowed for the convexity of the criterion) then the identification has been done with respect to the points C and P, giving a good compromise and leading to the constants $(I_{1m}=2.40$ MPa, $L_{s}=0,62)$. 
The Willam's criterion also makes use of the third invariant of the deviator $J'_3$. The identification with respect to the points C and T leads to the constant $e=0.5$ which is the limit for the convexity of the model (dashed curve in figure \ref{YSfig}); again, a better identification has been found from points C and P (plain curve in figure \ref{YSfig}), with the constants $(e=0.68,m_{o}=5.3$ MPa$)$.
It can be seen that these models (and many others, see \cite{ragueneau_07} for a more complete comparison) fail to describe precisely the particular shape of the concrete's yield surface. 


\section{The proposed criterion}
\label{main}
In compression, \emph{i.e.} when (every) $\si\leqslant0$, the yielding of concrete involves a diffuse microcracking, and high stresses. The mechanism in tension, \emph{i.e.} when (at least one) $\si\geqslant0$, is very different, with low stress levels and a quick localization of the microcracks \cite{fokwa_92}. A part of the present criterion is related to the von Mises norm of the stress deviator that appears to be relevant to the compression states, in the low confinement range. The other part uses a stress tensor exponential (recalled in equation \ref{derexpo}) in order to describe the dramatic effect of the positiveness of a principal stress, in  the sense of the Rankine criterion, \emph{via} an euclidean norm ($\| \A \| = (A_{\ti\tj}A_{\ti\tj})^{1/2}$). The constant $\sy$ possesses a limiting role; the second constant $\so$ rules the dissymmetry of the yield surface and limits the numerical value of the exponential term. The term $\sqrt{3}$ makes the second member equal to zero when $\SIG=\tens{0}$.
\begin{equation}
f(\SIG) = ||\SIGD|| + \so \left( \left\| \exp \left( \frac{\SIG}{\so} \right) \right\| - \sqrt{3} \right) - \sy
\label{yieldsurf}
\end{equation}
Rewritten in term of principal stresses $(\si, \sii, \siii)$, the equation (\ref{eigenyieldsurf}) exhibits clearly the role of the second member that is positive if at least one principal stress is positive. One can remark that the numerical calculus of a tensor exponential does not require to diagonalize it (it is an integrated function in many softwares).
\begin{eqnarray}
\tilde{f}(\si,\sii,\siii) &=& \sqrt{ \frac{(\sii-\siii)^2+(\siii-\si)^2+(\si-\sii)^2} {3} }\nonumber\\
&+& \so \left( \sqrt{ \exp \left(\frac{2 \si}{\so}\right) + \exp \left(\frac{2 \sii}{\so}\right) + \exp \left(\frac{2 \siii}{\so}\right) } - \sqrt{3} \right) - \sy
\label{eigenyieldsurf}
\end{eqnarray}
\begin{figure}[t]
\begin{center}
\includegraphics[scale=0.5]{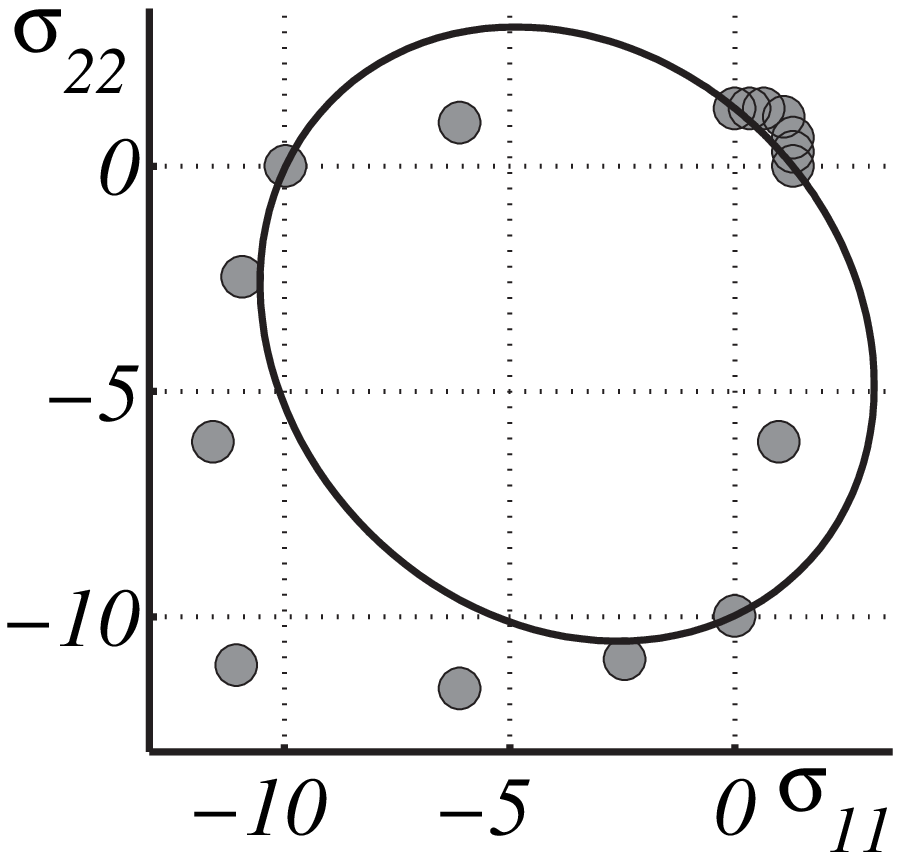}(a)
\includegraphics[scale= 0.5]{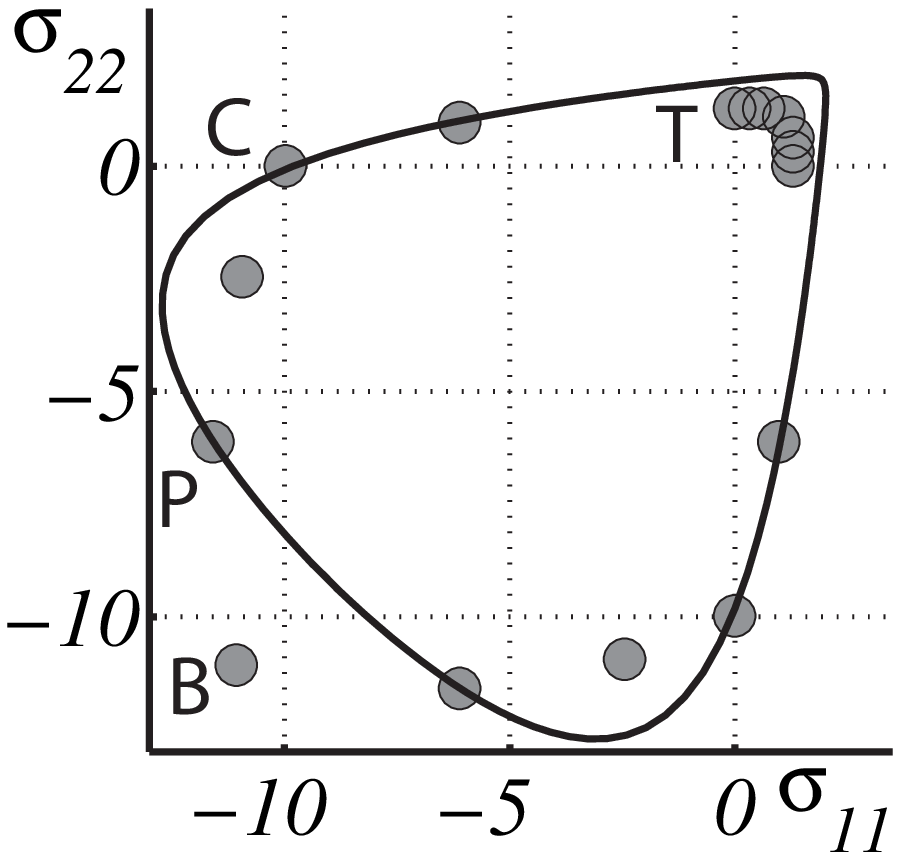}(b)\\
\includegraphics[scale= 0.5]{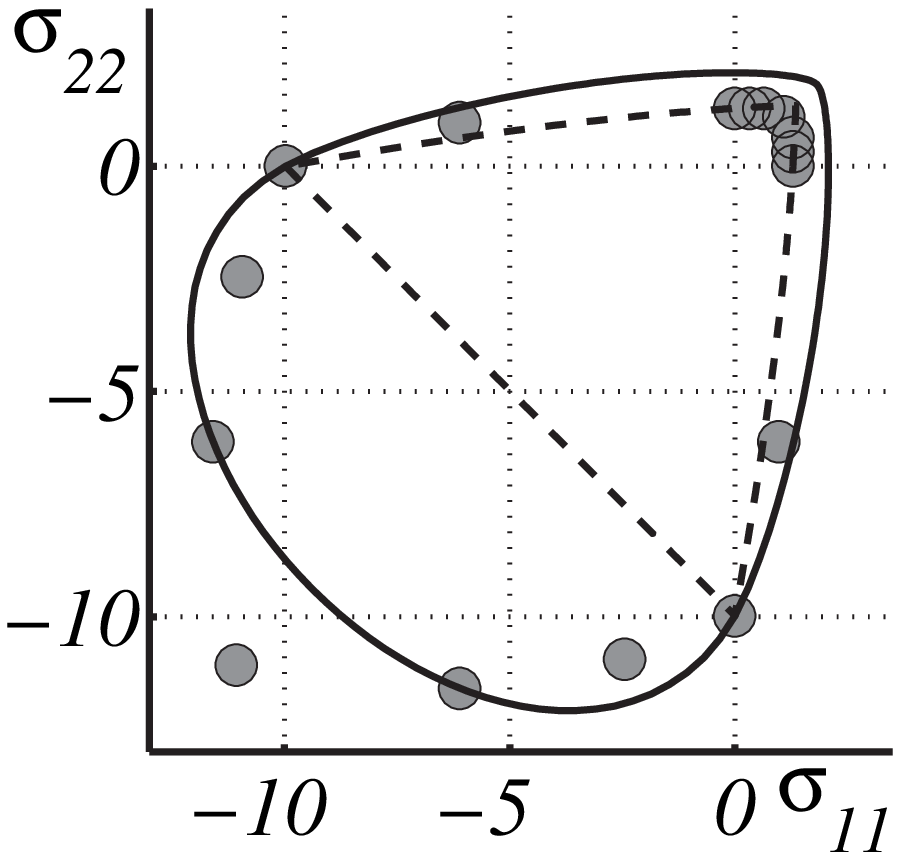}(c)
\includegraphics[scale= 0.5]{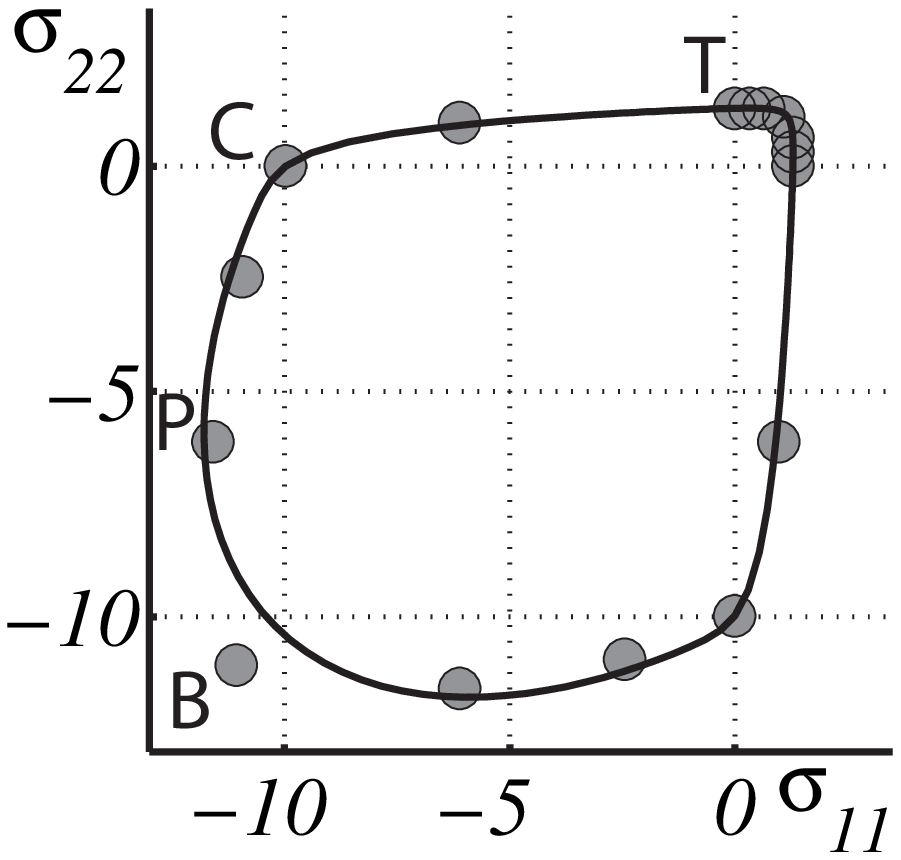}(d)
\caption{Yield surfaces (MPa) of De Vree (a), Ma\"iolino (b), Willam (c) and the present one (d). Circular dots represent Kupfer's data.}
\label{YSfig}
\end{center}
\end{figure}
The constants $(\so,\sy)$ can be identified with respect to the classical (and simplest) uniaxial compression and tension tests. Calling $\sit>0$ and $\sic<0$ the experimental elastic limit stresses in pure tension and compression, corresponding respectively to points T and C on figure (\ref{YSfig}), the difference $\tilde{f}(\sit,0,0)-\tilde{f}(\sic,0,0)=0$ leads to:\\

\begin{figure}[htbp]
	\begin{minipage}[c]{.48\linewidth}
		\centering
		\includegraphics[scale=0.18]{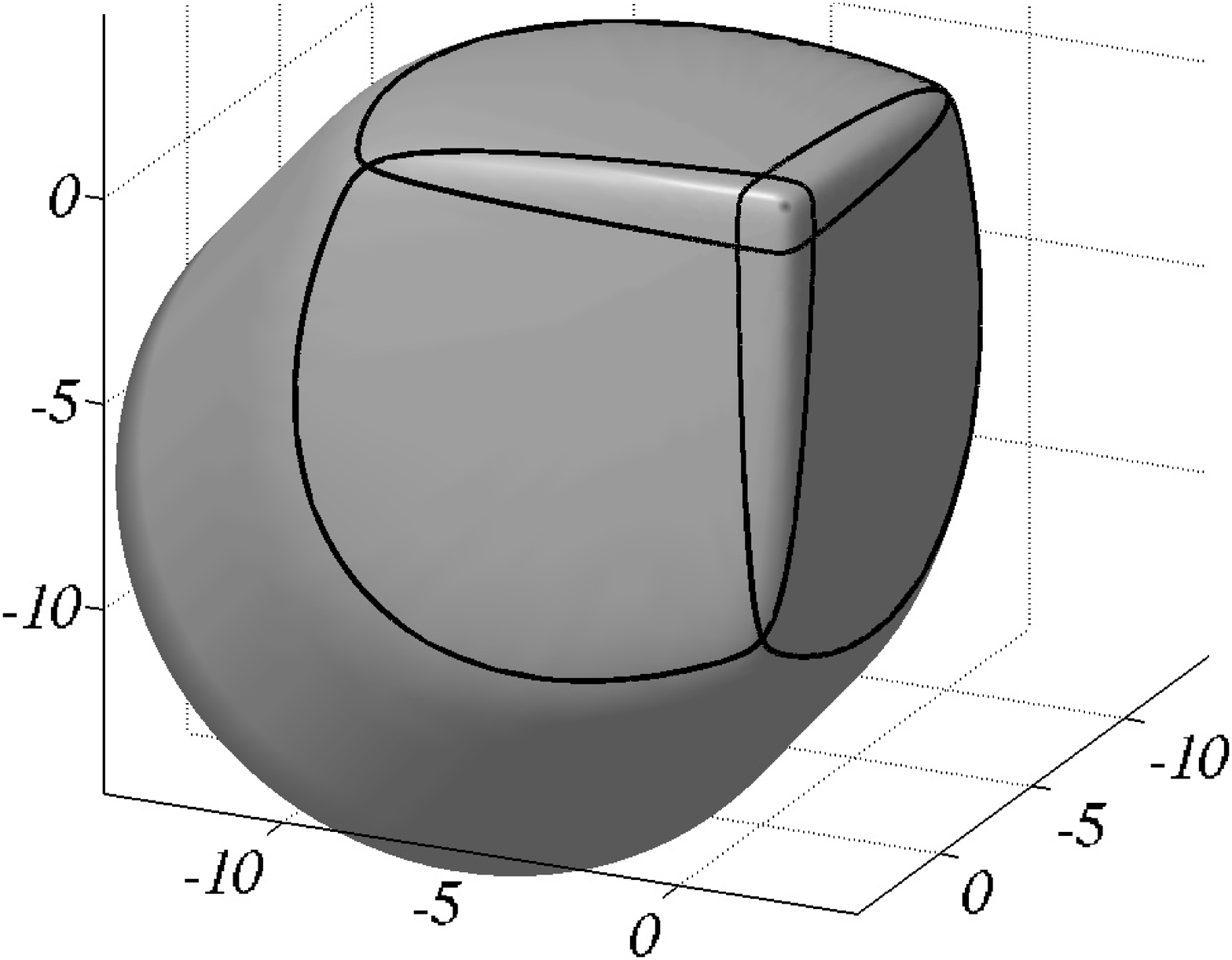}
		\caption{Yield surface in principal stresses space (MPa). The plain lines correspond to the intersections with planes $\sigma_{I}=0$.}
		\label{SSTD}
	\end{minipage} \hfill
	\begin{minipage}[c]{.48\linewidth}
		\centering
		\includegraphics[scale=0.5]{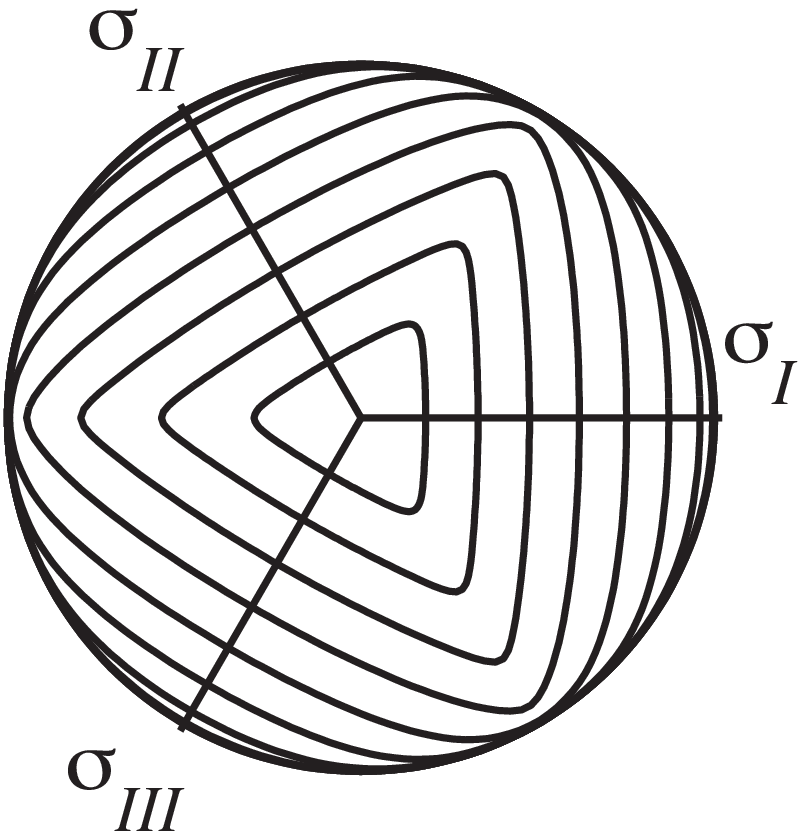}
		\caption{Deviatoric view of the criterion.}
		\label{SS_PlanHyd}
	\end{minipage} \hfill
\end{figure}

$\tilde{f}(\sit,0,0)-\tilde{f}(\sic,0,0)=0$

\begin{equation}
\so\sqrt{\exp\left(\frac{2\sit}{\so}\right)+2}+\sit\sqrt{\frac{2}{3}}=\so\sqrt{\exp\left(\frac{2\sic}{\so}\right)+2}-\sic\sqrt{\frac{2}{3}}\label{equiden}
\end{equation}
Each member is a monotonic function of $\so$: their unique intersection, numerically calculated, gives the value of $\so$. The equation $\tilde{f}(\sic,0,0)=0$, or $\tilde{f}(\sit,0,0)=0$, gives the second constant $\sy$. Here, the identified values are $\so=0,455$ MPa and $\sy=8,00$ MPa. The figure (\ref{YSfig}) shows the good agreement of this criterion with the biaxial testings. The strong curvature around the point C of simple compression is well depicted. The figure (\ref{SSTD}), in the principal stresses space, shows that this criterion can be seen as a softened version of the combination of von Mises (for the cylindrical shape) and Rankine (for flattened faces) ones. The figure (\ref{SS_PlanHyd})  shows the deviatoric sections of the criterion that evolve from the triangular shape of Rankine envelope to the circular shape of von Mises.

\section{Gradient, hessian and convexity}
\label{deriv}
The use of a yield surface in a damage (or plasticity) model is helped by the knowledge of its gradient. We can already calculate the gradient of $f$ with respect to the stress tensor $\SIG$. The derivative of the first member of equation (\ref{yieldsurf}) is straightforward. For the second one, it is useful to use the property (\ref{theo2}). The obtained gradient presents a simple (and intrinsic) expression:
\begin{equation}
\frac{\partial f}{\partial \SIG}=\frac{\SIGD}{\| \SIGD \|} + 
\frac{\exp(2\SIG/\so)}{\| \exp(\SIG/\so) \|}\label{gradf}
\end{equation}
The hessian of $f$ may also be useful in numerical calculus. For the first member of equation (\ref{yieldsurf}) we obtain the following expression, in which $\otimes$ represents the dyadic product and $\PD$ is the fourth rank projector onto the deviatoric subspace:
\begin{equation}
\frac{\partial^2 \| \SIGD \|}{\partial \SIG^2} = \frac{1}{\|\SIGD\|} \left( \PD - \frac{\SIGD}{\| \SIGD \|}\otimes \frac{\SIGD}{\| \SIGD \|} \right)
\textrm{ with }
P^{\mathrm{d}}_{\ti\tj\tk\tl} = \frac{1}{2}(\delta_{\ti\tk}\delta_{\tj\tl}+\delta_{\ti\tl}\delta_{\tj\tk}) - \frac{1}{3}\delta_{\ti\tj}\delta_{\tk\tl}
\label{hess1}
\end{equation}
The fourth rank operator between parenthesis in the equation (\ref{hess1}) is a projector onto the four dimensions subspace of the second rank symmetric and deviatoric tensors orthogonal to $\SIGD$; its eigenvalues are then $(1,1,1,1,0,0)$ which proves the well known convexity of the von Mises criterion. For the second member of equation (\ref{yieldsurf}), it is convenient to have a change of variable $\SIG/\so=\SIG'$:
\begin{equation}
\so \frac{\partial^2}{\partial \SIG^2}\left( \left\| \exp \left( \frac{\SIG}{\so} \right) \right\| \right) =
\frac{1}{\so} \frac{\partial^2 \| \exp(\SIG') \|}{\partial \SIG'^2}
\label{chandevar}
\end{equation}
Using the gradient (\ref{gradf}) leads to:
\begin{equation}
\frac{\partial^2 \| \exp(\SIG') \|}{\partial \SIG'^2} =
\frac{2}{\| \exp(\SIG') \|}\frac{\partial \exp(2\SIG')}{\partial (2\SIG')} - \frac{1}{\| \exp(\SIG') \|^3}\exp(2\SIG')\otimes\exp(2\SIG')
\label{dersec}
\end{equation}
Let us call $\vect{u}_\ti$ the eigenvectors and $\lambda_{\ti}$ the eigenvalues of $\SIG$. Then $\vect{u}_\ti$ are also the eigenvectors of $\SIG'$ and $\exp(\SIG')$ (these tensors are coaxial with $\SIG$) and $\lambda_{\ti}/\so$ are the eigenvalues of $\SIG'$. We consider now the orthonormal base of the second order symmetric tensors $\B_\tI$, with $I\in(1..6)$, \emph{i.e.} $\B_\tI:\B_\tJ=\delta_{\tI\tJ}$ (the symbol ":" represents the tensor contraction (inner product) defined, for any symmetric second order tensors $\A$ and $\A'$, as $\A:\A'= A_{\ti\tj}A'_{\ti\tj}$ and $\delta$ is the Kronecker delta).

\begin{eqnarray}
	\B_1 &=& \vect{u}_1 \otimes \vect{u}_1,\quad
	\B_4 = \left( \vect{u}_2 \otimes \vect{u}_3 + \vect{u}_3 \otimes \vect{u}_2 \right)/\sqrt{2}\nonumber\\
	\B_2 &=& \vect{u}_2 \otimes \vect{u}_2,\quad
	\B_5 = \left( \vect{u}_3 \otimes \vect{u}_1 + \vect{u}_1 \otimes \vect{u}_3 \right)/\sqrt{2}\nonumber\\
	\B_3 &=& \vect{u}_3 \otimes \vect{u}_3,\quad
	\B_6 = \left( \vect{u}_1 \otimes \vect{u}_2 + \vect{u}_2 \otimes \vect{u}_1 \right)/\sqrt{2}
\end{eqnarray}

From formula (\ref{derexpobp}), the second derivative (\ref{dersec}) expresses in the base $\B_\tI\otimes\B_\tJ$ (in the Voigt notation, the terms 44, 55 et 66 would be divided by 2), in the general case $(\lambda_{1}\neq\lambda_{2}\neq\lambda_{3})$, as:
\begin{equation}
\frac{\partial^2 \| \exp(\SIG') \|}{\partial \SIG'^2} = 
s^{-3/2}
\left[\begin{array}{cccccc}2p_{1} s-p_{1}^2 & -p_{1}p_{2} & -p_{1}p_{3} & 0 & 0 & 0 \\-p_{1}p_{2} & 2p_{2} s-p_{2}^2 & -p_{2}p_{3} & 0 & 0 & 0 \\-p_{1}p_{3} & -p_{2}p_{3} & 2p_{3} s-p_{3}^2 & 0 & 0 & 0 \\0 & 0 & 0 & r_{23} & 0 & 0 \\0 & 0 & 0 & 0 & r_{31} & 0 \\0 & 0 & 0 & 0 & 0 & r_{12}\end{array}\right]_{\B_I}
\textrm{ with }
\left\{
\begin{array}{c}
	r_{\mathrm{i}\mathrm{j}}=\so s\frac{\pei-\pej}{\lambda_{\mathrm{i}}-\lambda_\mathrm{j}}\\
	\pei = e^{2\frac{\lambda_{\mathrm{i}}}{\so}}\\
	s = \sum_{\mathrm{i}=1}^{3}\pei
\end{array}
\right.
\label{Hessieng}
\end{equation}
From the convexity of the exponential function, the $r_{\mathrm{i}\mathrm{j}}$ are positive if $\so$ is positive. The case $\lambda_{\ti}=\lambda_{\tj}$ is treated similarly from equation (\ref{derexpobp}) and it implies again $r_{\mathrm{i}\mathrm{j}}>0$. The bilinear form associated to this expression can be written as follows, where $\A$ is a general symmetric second order tensor:
\begin{eqnarray}
&&s^{3/2}\A.\left(\frac{\partial^2 \| \exp(\SIG') \|}{\partial \SIG'^2}\right).\A 
= p_{1}A_{11}^2+p_{2}A_{22}^2+p_{3}A_{33}^2+2 r_{23}A_{23}^2+2 r_{31}A_{31}^2+2 r_{12}A_{12}^2\nonumber\\
&&
+2p_{2}p_{3}(A_{22}^2-A_{22}A_{33}+A_{33}^2)
+2p_{3}p_{1}(A_{33}^2-A_{33}A_{11}+A_{11}^2)
+2p_{1}p_{2}(A_{11}^2-A_{11}A_{22}+A_{22}^2)\label{defpos}
\end{eqnarray}
This expression is obviously positive and, added with the convexity of the von Mises term of the criterion, it proves the (strict) convexity of the proposed criterion $f$. This result can also be obtained without the hessian: it is easy to prove that the second partial derivatives $\partial^2 \tilde{f}/\partial \si^2$ are positive thus, from the Ball's theorem (number 5.1) \cite{ball_77}, $f$ is convex.

\section{Conclusions}

The proposed elasticity criterion offers a good description of Kupfer's biaxial testings on concrete. It depends of two constants that can be easily identified with respect to uniaxial compression and tension curves. Its convexity is assured by a unique condition ($\so>0$), leading to mathematical robustness. Future work may consists in the development of a full damage model using this criterion as a yield surface. Another possible enhancement is to take the high confinement effects into account by replacing the von Mises part of the criterion by, for example, the Drucker-Prager or an elliptic closed expression.

\appendix

\section{Mathematics}
\label{maths}

Let $\A=A_{\ti\tj}\vect{e}_{\ti}\otimes\vect{e}_{\tj}$ be a symmetric second order tensor expressed in an orthonormal basis $\{\vect{e}_{\ti}\}$; indexes $(\ti,\tj,\tk,\tl)$ refer to it. Let $\lambda_{\tp}$ be the eigenvalues and $\{\vect{v}_{\tp}\}$ the orthonormal basis of eigenvectors of $\A$; indexes $(\tp,\tq,\tr,\ts)$ refer to it. The derivative of the power low $\A^n$ expresses as:
\begin{equation}
\left. \frac{\partial \A^n}{\partial \A}\right|_{\ti\tj\tk\tl} = \sum_{p=0}^{n-1}A^p_{\ti\tk}A^{n-p-1}_{\tj\tl}
\Rightarrow
\left. \frac{\partial \A^n}{\partial \A}\right|_\mathrm{pqrs}=
\left\{
\begin{array}{ll}
n \lambda_{\tp}^{n-1} & \mathrm{if~} (\tp=\tq=\tr=\ts) \nonumber\\
n \lambda_{\tp}^{n-1} & \mathrm{if~} (\tp=\tr)\neq(\tq=\ts) \mathrm{~et~}\lambda_{\tp}=\lambda_{\tq}\nonumber\\
\frac{\lambda_{\tp}^{n}-\lambda_{\tq}^{n}}{\lambda_{\tp}-\lambda_{\tq}} & \mathrm{if~} (\tp=\tr)\neq(\tq=\ts) \mathrm{~et~}\lambda_{\tp}\neq\lambda_{\tq}\\
0 & \textrm{else}\nonumber
\end{array}
\right.
\end{equation}
This allows to express the tensor exponential derivative as:
\begin{equation}
\exp(\A) = \sum_{n=0}^\infty \frac{\A^n}{n!}
\Rightarrow
\left.\frac{\partial \exp(\A)}{\partial\A}\right|_{\ti\tj\tk\tl} = \sum_{n=1}^{\infty}\frac{1}{n!}\sum_{p=0}^{n-1}A^p_{\ti\tk}A^{n-p-1}_{\tj\tl}\label{derexpo}
\end{equation}
Two interesting properties are obtained from the (double) contraction of the previous expression of the exponential with $\A$ on the right side (\emph{i.e.} $A_{\tk\tl}\,\partial (\exp(\A))_{\ti\tj}/\partial A_{\tk\tl}$) and with $\exp(\A)$ on the left side (\emph{i.e.} $(\exp(\A))_{\ti\tj}\,\partial (\exp(\A))_{\ti\tj}/\partial A_{\tk\tl}$).
\begin{equation}
\frac{\partial \exp(\A)}{\partial\A}: \A = \A.\exp(\A)\textrm{~;~}\quad
\exp(\A): \frac{\partial \exp(\A)}{\partial\A} = \exp(2 \A)\label{theo2}
\end{equation}
In the base of eigenvectors $(\vect{v}_{\tp}\otimes\vect{v}_{\tq}\otimes\vect{v}_{\tr}\otimes\vect{v}_{\ts})$, the derivative (\ref{derexpo}) has the following components:
\begin{equation}
\left. \frac{\partial \exp(\A)}{\partial \A}\right|_\mathrm{pqrs}\left\{
\begin{array}{ll}
\exp(\lambda_{\tp}) & \mathrm{if~} (\tp=\tq=\tr=\ts)\nonumber\\
\exp(\lambda_{\tp}) & \mathrm{if~} (\tp=\tr)\neq(\tq=\ts) \mathrm{~et~}\lambda_{\tp}=\lambda_{\tq}\nonumber\\
\frac{\exp(\lambda_{\tp})-\exp(\lambda_{\tq})}{\lambda_{\tp}-\lambda_{\tq}} & \mathrm{if~} (\tp=\tr)\neq(\tq=\ts) \mathrm{~et~}\lambda_{\tp}\neq\lambda_{\tq}\label{derexpobp}\\
0 & \textrm{else}\nonumber
\end{array}
\right.
\end{equation}


\end{document}